\documentclass{PoS}
\usepackage{units}
\usepackage{comment}
\usepackage{amsmath}
\usepackage{mathtools}
\usepackage{ae, aecompl}
\usepackage{listings}
\usepackage[binary-units]{siunitx}
\usepackage{rotating}
\usepackage{subfig}
\usepackage{mparhack}

\title{Silicon Photomultipliers for Orbital Ultra High Energy Cosmic Ray Observation}

\ShortTitle{SiECA}

\author{\speaker{W.~Painter}$^2$, A.~Haungs$^1$, T.~Huber$^{2,4}$, M.~Karus$^1$, A.~Menshikov$^3$, M.~Oehler$^2$, M.~Renschler$^2$ - JEM-EUSO Collaboration\footnote{for collaboration list see PoS(ICRC2019)1177} \\
\llap{$^1$} Institut f\"ur Kernphysik, KIT - Karlsruhe Institute of Technology, Germany\\
\llap{$^2$} Institut f\"ur Experimentelle Teilchenphysik, KIT - Karlsruhe Institute of Technology, Germany\\
\llap{$^3$} Institut f\"{u}r Prozessdatenverarbeitung und Elektronik, KIT - Karlsruhe Institute of Technology, Germany\\
\llap{$^4$} DESY, Germany\\
E-mail: \email{andreas.haungs@kit.edu}, \email{william.painter@kit.edu}}

\abstract{Development of the Silicon photomultiplier Elementary Cell Add-on camera (SiECA) has provided extensive information regarding the use of SiPMs for future cosmic ray detection systems. We present the technical aspects of sensor readout development utilizing Citiroc ASIC chips from Weeroc controlled by a Xilinx FPGA to process and package events from four 64 channel Hamamatsu MPPC S13361 arrays generating 128 frame events with an integration time of $2.5\,\mu$s (parameters are based on JEM-EUSO geometry but can be easily adjusted). With single photon counting capability, SiECA proves SiPM are viable sensors to replace Multi-Anode PhotoMultiplier Tubes in future devices, especially when high luminosity exposure is possible potentially damaging MAPMT based systems. Complementary to the technical aspects, computational and analysis methods for sensor array characterization and in depth device flat-fielding are presented. Provided channel by channel biasing, in comparison to uniform biasing with MAPMTs, fine tuning of operating parameters with MPPC arrays allows for substantial improvements in detector and signal uniformity. 
}

\FullConference{36th International Cosmic Ray Conference -ICRC2019-\\
		July 24th - August 1st, 2019\\
		Madison, WI, U.S.A.}

\begin{document}

\setcounter{page}{2}

\section{Introduction}
This Paper describes the construction and characterization of SiECA, the Silicon  photomultiplier Elementary Cell Add~on camera for  or use in the detection of the fluorescence signature  of extensive air-showers generated by ultra-high energy cosmic rays in Earth's atmosphere. This work is predominantly  technical and hardware oriented with the physics motivation of operation of SiPMs as part of the orbital cosmic ray observatory in the Extreme Universe Space Observatory (EUSO~\cite{EUSO-Program}) or Probe Of Extreme Multi-Messenger Astrophysics (POEMMA~\cite{poemma}) telescopes (see also the R\&D studies for the use of SiPM in EUSO-SPB2 and POEMMA~\cite{otte_icrc19,osteria_icrc19}).

The evaluation of SiPMs as photon-sensors for a non-terrestrial cosmic ray detecting telescope, or collection 
of telescopes, is the goal of the SiECA R\&D work~\cite{sieca-icrc2017}. 
Designed as an ancillary device to be easily attached to existing fluorescence telescopes, 
utilizing the same optics, internal voltage generation from a 5V supply, internal or external clock 
and external trigger, serial USB data and communication line, SiECA can be easily mounted and 
evaluated at any operational cosmic ray telescope provided available space. 
The development of SiECA from concept to first deployment on-board the EUSO-SPB1~\cite{SPB-Flight} balloon borne telescope was a mere 18 months. Many improvements remain to be made, however, substantial understanding of the SiPM and ASIC characteristics have been gained in this process.

First deployment tests occurred with the launch of the 2017 NASA Super-Pressure Balloon mission carrying the JEM-EUSO prototype telescope, EUSO-SPB. SiECA was attached to the side of the Photo-Detection Module 
(PDM, EUSO-SPB's main detector) so as to be in the same focal plane and make use of the optics constructed for the mission.
As SiECA is designed to evaluate the capability of MPPC for cosmic ray detection, this parallel testing campaign provides direct comparison of MPPC and MAPMT under identical measurement conditions. Due to the complications of the 2017 NASA Super Pressure Balloon campaign, search for cosmic ray and other interesting signatures with SiECA was not possible. After the loss of SiECA in the deep ocean, a replica could be produced at KIT, with which further test and calibration measurements are performed. Despite the less than optimal deployment test on-board EUSO-SPB1 (the balloon had to be released after 12 days of flight with only few hours of SiECA measurements), significant advancement has been made with respect to the hardware implementation. The first part of this contribution is devoted to the hardware of SiECA and the second part to its laboratory calibration. Both were necessary in the creation of a working detector.

\section{SiECA}
SiECA is a 256 channel photon camera constructed of three printed circuit boards: the Elementary Cell (EC) 
front board supporting the four 64 channel Multi-Pixel Photon Counters (MPPC or SiPM) sensors (Hamamatsu 
S13361-3050AS-08), and the two readout boards called the Data Acquisition (DAQ) and Mezzanine 
(Mezz) boards housing the eight ASIC (Weeroc Citiroc 1A) chips, FPGA (Spartan 6 
XC6SLX100-2FGG676I), eight bias voltage generators (Hamamatsu C11204-02), interface control 
chipsets (USB and LVDS) and other required analog and digital logic components.

\subsection{SiECA Mechanical Design}

The physical arrangement of the sensors is well established by the EUSO PDM construction. The necessary space allotment for the bias voltage generation, MPPC sensor connection, thermal sensors, ASIC, FPGA, supply voltage regulation, USB and LVDS interface chips have been determined based on the sensor placement and attempting to follow the PDM structure. 
Due to the short development period cumulating in the EUSO-SPB1 launch, a simple design was adopted utilizing an elongated Si-EC board holding the sensors and connected to the readout boards oriented perpendicular behind the Si-EC board. The boards are mechanically supported by rigid frames, originally in aluminum but, rebuilt with 3D printed plastic components for electrical isolation and minimal production costs and time. 
The mechanical design of SiECA is driven by the electrical requirements, aside from the sensor orientation and proximity to the edge of the camera for alignment. 
\begin{figure}[t]
 \centering
 \includegraphics[width = 0.3\textwidth]{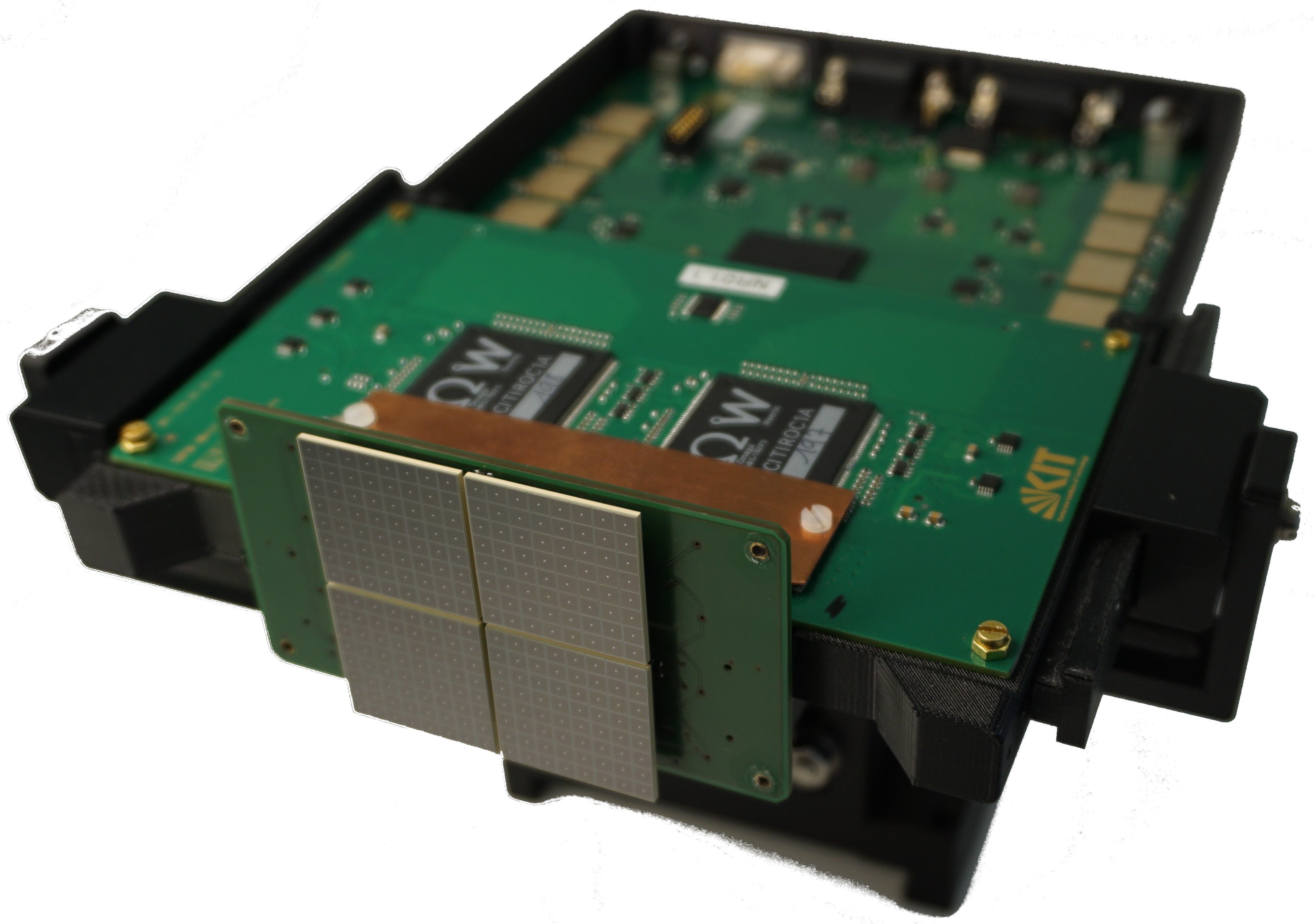} \hspace*{0.1cm}
 \includegraphics[width = 0.65\textwidth]{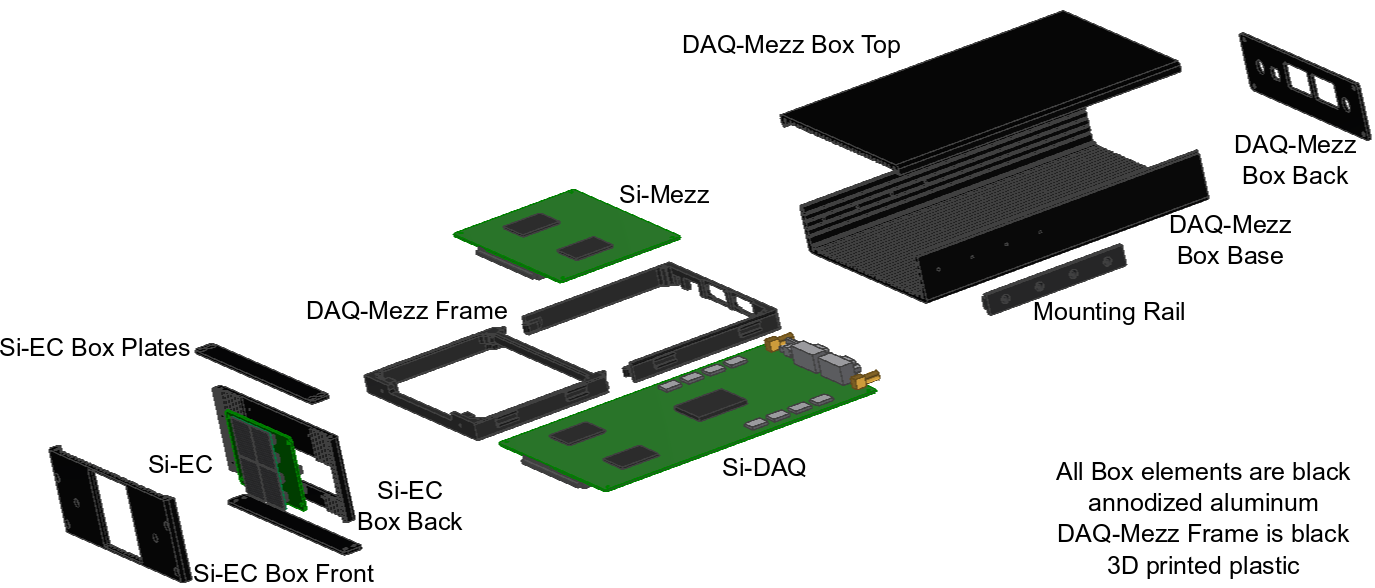}
\caption{Left: SiECA Assembled; Right: SiECA complete with isolating aluminum black box and mounting rail.} \label{fig:SiECA-Complete} 
\end{figure}

\paragraph{Si-EC board}
The board connects the four S13361 arrays, distributes bias voltages and collects measurement signals from each MPPC channel. Additionally, four temperature sensors located between the corners of neighboring MPPC arrays monitor and report the operating temperature of the sensors to the FPGA. While not implemented in the EUSO-SPB1 deployment, these temperature sensors could be used to regulate the bias voltage generators to maintain constant internal gain of the sensors. 
The board's dimension is \SI{90.6 x 50.6 x 1.6}{mm} and consists of eight conductive layers, four for bias voltages and signals and four for isolating ground planes for reduced crosstalk between signal traces. Four  mounting holes are located in the corners for securing the Si-EC board to the support structure to prevent mechanical loading of the electrical connectors. 
The Si-EC board connects perpendicularly to the readout boards so as to minimize the occupied area parallel to the focal surface and to be similar in structure to the PDM. 

\paragraph{Mezzanine Board}
The \SI{91 x 150}{mm} Mezzanine board houses four of the Citiroc ASIC and all the passive components necessary for processing half of the MPPC signals generated on the Si-EC board. Connection to the DAQ for initialization and measurement signal transmission is made via  connector pairs located behind and outward from the ASIC. 

\paragraph{DAQ Board}
The DAQ board is the core of SiECA. From the front,  64 MPPC channels being distributed, on either side of the board, to the two 32 channel Citiroc ASIC. A small ADC for converting the temperature sensor measurements to digital values for the FPGA is also included. The central chip is the Spartan6 FPGA which controls initialization, trigger counting, event packet generation and packet distribution to the USB interface.

\subsection{SiECA Electrical Design}

\paragraph{Si-EC Board}
The mechanical properties and positioning of the sensor connectors on the Si-EC board are dictated by the spacing and orientation of the MPPC. 
To minimize costs, through-vias are used throughout the board.  
All the passive elements are rated for \SI{100}{\volt} even though operating bias voltages are below \SI{60}{\volt} for the MPPC.  
To unify the grounding planes, stitching vias have been added where possible. This is easiest around the outer area of the board but efforts were made to include them in the more densely traced inner areas. This could be further optimized given additional development time, however, better assessment of the noise impact on the signal quality is necessary to warrant additional grounding.  

\paragraph{Data Acquisition and Mezzanine Boards}
The eight bias voltages, tuned for the MPPC channels they supply, are generated by the eight C11204-0 chips. 
They are connected to the Si-EC board by the outermost eight channels (two on the top and bottom of each end) of the DAQ board. One channel on the top and bottom is left non-connected followed by the analog ground. The next two pins inward contain the \SI{3.3}{\volt} supply and signal return connections for each of the four temperature sensors on the Si-EC board. The Mezz connection to the Si-EC board is the same as the DAQ however the Mezz connection does not carry any bias voltages or temperature signals leaving these pins non-connected.
The DAQ and Mezz boards contain the ASIC, FPGA, bias voltage generators, and interface chipsets. Incoming MPPC signals are terminated with \SI{50}{\ohm} resistors and \SI{100}{\nano\farad} capacitor in series to ground. Combined, the two boards contain eight Citiroc capable of analyzing 256 continuous signals from the MPPC front board. 
The output of these ASICs is 256 digital trigger lines, 16 multiplexed charge readouts (one high and one low gain per chip). 
The digital trigger lines are individually connected to the FPGA for trigger counting and the multiplexed lines are connected for charge measurement. The Mezz board signals are connected to the DAQ board. These connectors also supply power, initialization and clocks to the Mezz Citiroc. 
The analog temperature measurement is run through the ADC located centrally, between the Citiroc and FPGA on the DAQ before being connected to the FPGA where averaging across two neighboring temperature sensors provides an estimated temperature for each MPPC array.

\section{SiECA Signal Processing and Data Generation}

\paragraph{Initialization of System}
When fully connected and powered, SiECA will automatically reload the operation settings from the previous operation period.  Depending on the state of the flags  the previous bias voltages and voltage offsets will be reset from the provided configuration file. Parameters in the ASIC such as pre-amplifier gains, discriminator thresholds and fine tunings, and FPGA parameters such as event depth, GTU Gate Time Unit) length, and trigger latency are set by additional flags. Once the system is operational and the current consumption from initializing all the on-board hardware has dropped to the stable, low level, the event ring buffer is being filled with the specified measurements from each channel every GTU. When a trigger is received, the specified number of consecutive GTU buffer bins is read out with the defined latency. 

 \paragraph{Sensor and Front Board Signal Handling}
The MPPC used in SiECA are arrays of parallel reverse biased GAPD (Geiger mode Avalanche Photo Diode). In order for an electronic signal to be generated when a photon interacts with a GAPD, the biasing voltage must be above the breakdown voltage of the GAPD. To ensure constant and stable biasing voltages, each of the eight bias voltage generator traces is decoupled to ground.  
Post routing assessment shows the range of signal trace lengths span from \SIrange{6}{25}{\milli\metre}. Concern over the signal trace length is confined to the Si-EC board as the highly parallel distribution from the connectors to the ASIC on the DAQ and Mezz boards are uniform and with little deviation. 
Given the intended integration time of \SI{2.5}{\micro\second}, the relative difference in signal arrival timing between two coincident photons is negligible. 
Substantial noise was generated by the temperature sensors when operated in parallel with the MPPC. This was corrected in the FPGA firmware causing the temperature to be measured only at start-up and after an event is triggered, meaning the bias voltage was not regulated by the temperature sensors. Connection of the temperature sensors with a dedicated grounding scheme rather than using the same ground as the MPPC could reduce the impact of measurement at the same time at the cost of a more complicated PCB layout.

\paragraph{DAQ and Mezz Digitization with Citiroc}
In the design of the SiECA camera, we selected components that best embodied the operational goals of the EUSO telescope design. Consideration of several ASIC was undertaken early in the design process. The selection of the Citiroc ASIC was based primarily on design comparability with the MAPMT ASIC, the Spaciroc.  The ability of Citiroc to function in a peak over threshold or charge integration mode proved critical in the design of  SiECA despite the charge integration readout time being far too slow for EUSO-like measurements and certainly too slow for Cerenkov events. 
 Signals are amplified with either low or high gain pre-amplifiers. These increase the voltage of the signal allowing for the shapers to properly distinguish peaks from baseline noise.  Photon counting loses resolution on multiple incident photons at a time but the triggers are processed by the FPGA continuously so dead-time is negligible between GTU. 

\paragraph{FPGA Gate Timing, Trigger Processing and Packet Generation}
The FPGA handles initialization, operation, timing, peak counting, trigger processing and event packet generation. SiECA is designed to accept external clock signals. In the absence of these signals, they can be internally generated with local oscillators and synchronized with the hosts GPS PPS signal.  After initialization, SiECA will begin writing the count of pulses above threshold for each channel, each GTU, into the ring buffer. This buffer is capable of simultaneously storing the last 1024 GTU corresponding to \SI{2.56}{\milli\second}. This measurement is continuous, overwriting the oldest bin with the newest by reference such that at any time a trigger can be received and corresponding GTU bins can be collected, wrapped with header and footer information, and pushed to the FIFO buffer.

\section{Camera Calibration}
The calibration of the SiECA camera, and any multi-sensor detector, is carried out so as to minimize the post processing and artificial correction needed to extract physically relevant measurements. In SiECA, this is primarily handled by the flat fielding of the camera. Flat fielding is the process by which a set of sensors are tuned so that for an equal signal, equal responses are generated by each element. The nature of the response is dependent on the tuning, in this case, tuning for uniform gain on each sensor does not completely unify the photon detection efficiency (PDE).  The remaining non-uniformity must be corrected in post processing to accurately reconstruct any measured event.  
Measurements  have come from the single channel readout board~\cite{MPPC64Calib-Max}. Application of these measurements allows for tuning of the bias voltage applied to each MPPC channel through the input DAC of the Citiroc to achieve a camera with a focal surface uniform in either gain or PDE.

\paragraph{Single Channel Illumination}
To determine the necessary ASIC settings, specifically the gain and threshold for each channel, to operate in peak counting mode, a parameter scan has been developed. By scanning over the gain and threshold combinations for a set of voltages, it is possible to see the necessary threshold setting for a given gain and bias voltage corresponding to the discrete photo-electron (PE) amplitudes.  In this measurement the light intensity is set to three photons per pulse with a pulse frequency of \SI{1}{\mega\hertz}. This frequency ensures that each \SI{2.5}{\micro\second}-GTU is illuminated on average 2.5 times or an average of 7.5 photons will be incident on the illuminated channel each GTU.  One hundred events, each containing \SI{128}{GTU}, are recorded before moving to the next settings. A script checks the file size for each set of parameters and if it does not match what is expected, the file is rerun. 
\begin{figure}[ht]
  \centering
  \includegraphics[width=0.54\textwidth]{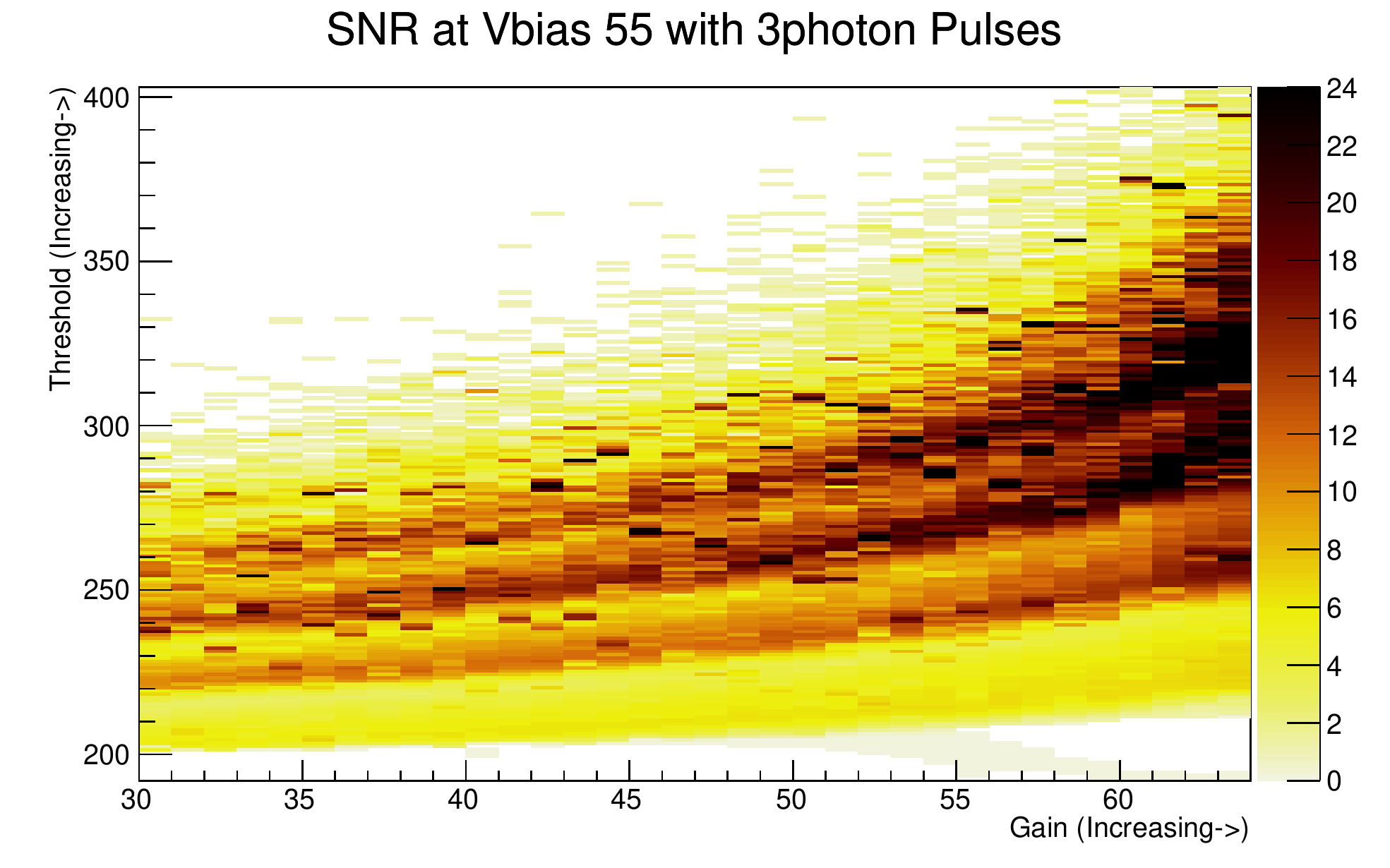} \hspace*{0.1cm}
	\includegraphics[width=0.35\textwidth]{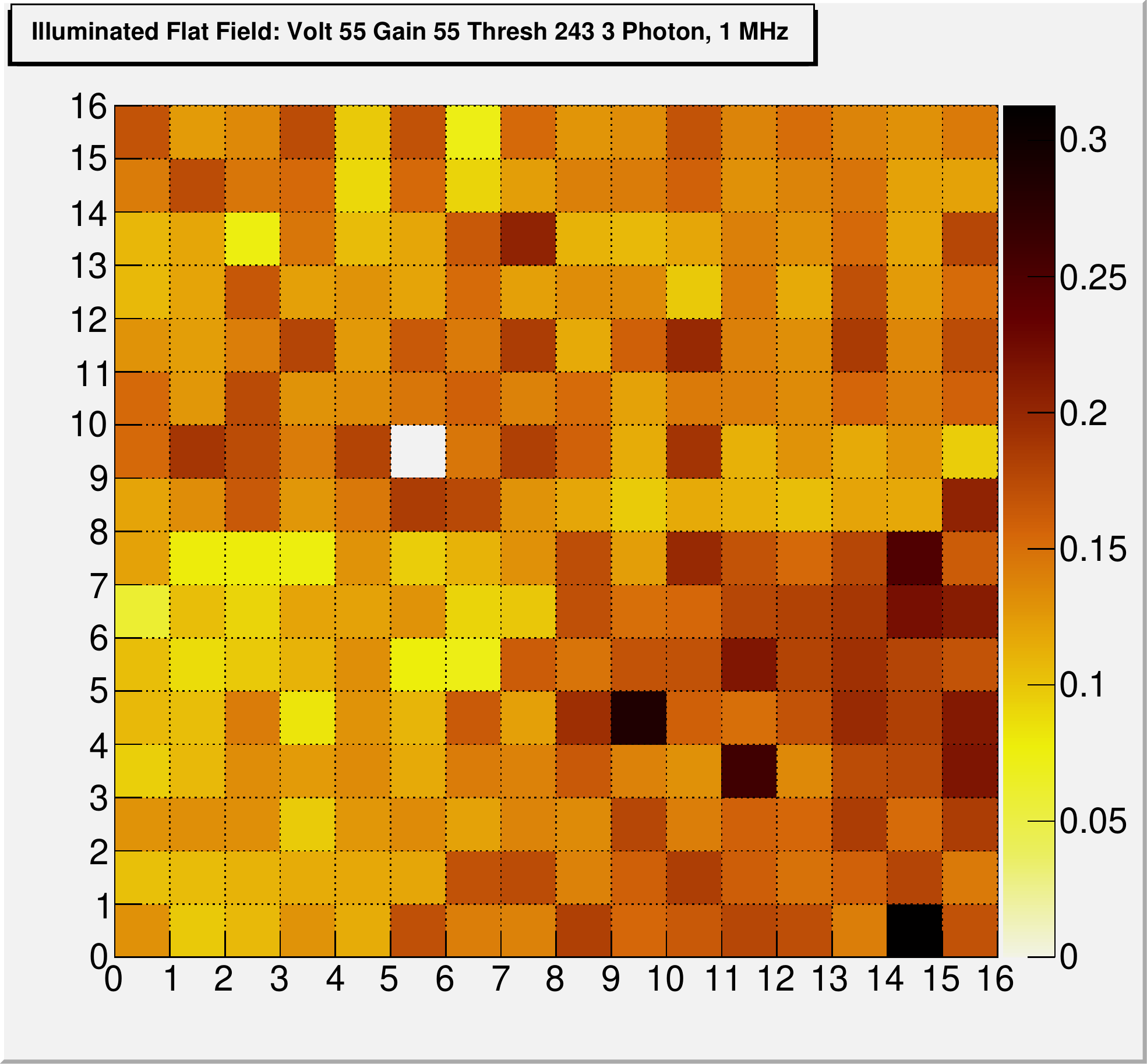}
  \caption{Left: Signal-to-Noise ($SNR$) ratio plot for $V_{bias}$ = \SI{55}{V}; Right: Example flat field results with EUSO-SPB1 flight parameters}
  \label{fig:V553photon}
\end{figure} 

Figure~\ref{fig:V553photon}, left panel, is instructive in illustrating the distinct separation of PE peaks within the measurement parameter space. To understand the implications, we look at the shape for one gain. Moving upward in increasing threshold from the X-axis we start in a null space in which the threshold is within the 0-PE pedestal. As the pedestal is uniform and measured by all channels, illuminated or not, whenever a PE cascade is not present, the illuminated channel sees less pedestal events than the average dark channel. 
For this plot, all negative signal-to-noise ($SNR$) values have been set to 0 to avoid confusion. 
Above the pedestal, we enter the 1-PE peak signal amplitude. 
This region is notably lighter than the multi-PE peak due to the signal generated by drifting electrons in the depletion region of the GAPD in each MPPC channel. For the room temperature measurement, \SI{0.5}{\mega\hertz} dark count rate is expected which is of similar magnitude to the \SI{7.5}{\mega\hertz-photon} signal. Above this level, dark noise is increasingly suppressed as multi-PE dark counts are energetically unlikely. The next dark band is the 2-PE amplitude threshold which is notably darker due to the substantially decreased dark count rate achieving this charge amplitude. Subsequently higher PE peaks are distinct for low to moderate gains but become indistinct due to the non-linearity in the amplifier at high gain levels. 

From this plot, determination of the threshold and gain setting can be selectively determined for exclusion of the pedestal and dark count events but also for higher thresholds limiting the selection to several-PE signals at the cost of decreased collection and energy resolution. For operation, combining the measured background signal and the laboratory reference plots, the voltage, gain and threshold necessary can be determined by inspection. 

\paragraph{Illuminated Flat Field test}
Verification of the calculated flat field corrections made in the bias voltage generators and ASIC can be verified by the use of either full field illumination in which the entire focal surface is uniformly illuminated or calibrated measurement at each channel. The flat focal surface of SiECA contains 256 MPPC channels, each with individual parameters. By flat fielding the sensors in gain, the measured signal amplitude for a single GAPD discharge should be the same regardless of channel. Thus, measurement of the uniformity after flat field corrections provide information about the parameters not flat fielded upon, PDE in the case of a gain flat field, as well as systematic issues such as electronic cross-talk due to PCB layout. 

Since full events are recorded, with every channel's count in each GTU included, we can make assessment of the uniformity under illumination but also the impact of illumination on a neighboring channel or electronic noise. Thus, we have insight into both the uniformity of each sensor and the electronic effects of the supporting electronics. This type of scan will also highlight any channel mapping errors such as a mismatch of ASIC orientation in the hardware and analysis software. For simplicity, we will look at uniformity, the number of counts on each channel per GTU of illumination and the noise impact in neighboring channels of the illuminated channel. In this instance, neighboring is taken to be the up to eight channels (four meeting at a common edge and four meeting at a common corner) closest to the illuminated channel though in principle the influence of any channel could be assessed on any other channel. 

As is apparent from Figure~\ref{fig:V553photon}, right panel, channel 3B3 is not functioning properly, indicated by a zero count rate. Otherwise, the uniformity is good, showing variation of less than one photon per GTU while still uncorrected for dark noise counts. A rough estimation given \SI{3}{photons/pulse} at \SI{1}{\mega\hertz} for \SI{2.5}{\micro\second} GTUs should see \SI{7.5}{photons/GTU} at the sensor. The lower count rate indicates a substantial loss of signal in processing.
Due to the variation in channel sensitivity, due to non-uniform PDE, this variation is expected. Correction by flat fielding on PDE, then correcting the gain non-uniformity with the Citiroc pre-amplifier, and if necessary a channel specific triggering threshold, should result in much more uniform sensitivity across the device. 

\paragraph{Summary Calibration} 
The application of characterization measurements for constructing uniform arrays of sensors, the following points are of the most significance:  (i) Flat fielding on Gain (similar amplitude signal; PDE must be corrected manually; better for charge integration when many photons are arriving); (ii) Flat fielding on PDE (uniform probability of detection across sensitive surface; gain and Threshold can be corrected in ASIC; better for individual photon counting); (iii) Full field flat fielding is more efficient; (iv) Channel by channel flat fielding indicates readout chain noise issues.
Once calibration is complete, the resulting camera is properly optimized for measurement, at least as well as is possible in the laboratory.

\section{Conclusion}
The development of the 256-channel SiPM-based SiECA camera and subsequent test flight expereiences as well as laboratory calibration efforts has provided extensive information about the operation of SiPM at low 
temperatures and pressures in the upper atmosphere. 
Generally, the use of silicon based photomultipliers for downward looking fluorescence and Cherenkov 
UHECR EAS signatures is promising and the benefits from use of a more robust, compact, 
low voltage device represent the future of non-terrestrial cosmic ray research.
The next step is the deployment of SiECA to a facility or installation which maximizes the possibility 
to achieve the scientific goals set out prior to the construction of the camera.

\end{document}